%% file: main.tex
\documentclass[conference]{IEEEtran}
\IEEEoverridecommandlockouts
\usepackage{pifont}

\usepackage{cite}
\usepackage{amsmath,amssymb,amsfonts}
\usepackage{algorithmic}
\usepackage{graphicx}
\usepackage{textcomp}
\usepackage{xcolor}

\usepackage{authblk}

\def\BibTeX{{\rm B\kern-.05em{\sc i\kern-.025em b}\kern-.08em
    T\kern-.1667em\lower.7ex\hbox{E}\kern-.125emX}}

\usepackage{blindtext}
\usepackage{subfigure}
\usepackage{verbatim}
\newcommand{\solution}{Rina}

\begin{document}

% \title{Conference Paper Title*\\

% for Incremental Deployment and Robustness 
\title{Rina: Enhancing Ring-AllReduce with In-network Aggregation in Distributed Model Training
\thanks{* Corresponding author: Yang Xu (xuy@fudan.edu.cn)}
\thanks{Zixuan Chen (zxchen20@fudan.edu.cn)}
\thanks{To appear in ICNP 2024. Preview version only.}
}

\input{Contents/Authors}

\maketitle

\input{Contents/Abstract}


\input{body}

\bibliographystyle{IEEEtran}
\bibliography{main}

\end{document}

%% file: Contents/Authors.tex
\author[$\dagger$]{Zixuan Chen}
\author[$\dagger$]{Xuandong Liu}
\author[$\dagger$]{Minglin Li}
\author[$\dagger$]{Yinfan Hu}
\author[$\dagger$]{Hao Mei} 
\author[$\dagger$]{\authorcr Huifeng Xing}
\author[$\dagger$]{Hao Wang}
\author[$\dagger$]{Wanxin Shi}
\author[$\dagger \ddagger$]{Sen Liu}
\author[$\dagger \diamondsuit \ast$]{Yang Xu}
\affil[$\dagger$]{School of Computer Science, Fudan University, Shanghai, China}
\affil[$\ddagger$]{Institute of Financial Technology, Fudan University, Shanghai, China}
\affil[$\diamondsuit$]{Pengcheng Laboratory, Shenzhen, China}

% \author{Anonymous Authors}
% \author{Paper \#309}
% \and
% \and
% \and
% \and
% \and

%% file: Contents/Abstract.tex
\begin{abstract}

% Distributed Deep Learning (DDL) has frequently been implemented to enhance the throughput of Deep Learning (DL) training. The two primary DDL synchronization structures utilized are Parameter Server (PS) and Ring-AllReduce (RAR). PS is known to struggle with severe many-to-one ``incast'' issues, subsequently causing considerable communication bottlenecks, while RAR is beset with long dependency chains. Both affect DDL throughput significantly. The introduction of In-network Aggregation (INA) has been instrumental in mitigating the throughput bottleneck issues prevalent in PS recently. 

Parameter Server (PS) and Ring-AllReduce (RAR) are two widely utilized synchronization architectures in multi-worker Deep Learning (DL), also referred to as Distributed Deep Learning (DDL). However, PS encounters challenges with the ``incast'' issue, while RAR struggles with problems caused by the long dependency chain. The emerging In-network Aggregation (INA) has been proposed to integrate with PS to mitigate its incast issue. However, such PS-based INA has poor incremental deployment abilities as it requires replacing all the switches to show significant performance improvement, which is not cost-effective. In this study, we present the incorporation of INA capabilities into RAR, called RAR with In-Network Aggregation (Rina), to tackle both the problems above. Rina features its agent-worker mechanism. When an INA-capable ToR switch is deployed, all workers in this rack run as one abstracted worker with the help of the agent, resulting in both excellent incremental deployment capabilities and better throughput. We conducted extensive testbed and simulation evaluations to substantiate the throughput advantages of Rina over existing DDL training synchronization structures. Compared with the state-of-the-art PS-based INA methods ATP, Rina can achieve more than 50\% throughput with the same hardware cost.

\end{abstract}

% incremental deployment issue and the inherent long dependency chain issue of RAR. 

%% file: body.tex
\section{Introduction}

The field of Deep Learning (DL) has seen remarkable advancements in recent years, driving transformative breakthroughs across various domains. The advent of Artificial General Intelligence (AGI) and large language generation models, such as the Generative Pre-trained Transformer (GPT)~\cite{openai2023gpt4}, have significantly advanced machine intelligence and Natural Language Processing (NLP)~\cite{vaswani2017attention,stahlberg2020neural}. Furthermore, models like Segment Anything (SA)~\cite{kirillov2023segment} have enriched DL's capability in image segmentation.

As model complexity and dataset sizes continue to expand exponentially, Distributed Deep Learning (DDL) has emerged as a pivotal approach for efficient training. Data parallelism, a strategy for managing vast datasets, divides the dataset among distinct processing units for concurrent training. This technique enables DL models to accommodate substantially larger datasets, thereby significantly enhancing training efficiency.

\begin{figure}[tbp]
    \centering
    \includegraphics[width=0.8\columnwidth]{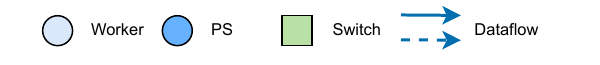}
    \subfigure[Parameter server]{\includegraphics[width=0.45\columnwidth]{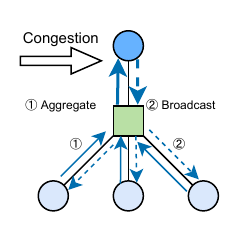}
    \label{fig:intro:ps-arch}}
    \subfigure[Ring-AllReduce]{\includegraphics[width=0.45\columnwidth]{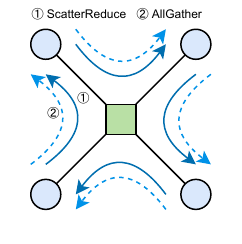}
    \label{fig:intro:rar-arch}}
    \caption{Popular synchronization architectures in DDL.}
    \label{fig:intro:sync_arch}
    \vspace{-0.2in}
\end{figure}

As depicted in Figure~\ref{fig:intro:sync_arch}, Parameter Server (PS) and Ring-AllReduce (RAR) are two prominent methods employed for parameter synchronization in data parallelism. PS operates by partitioning the dataset across multiple servers, thereby enabling parallel processing units to update these parameters independently. Conversely, RAR adopts a decentralized approach wherein each processing unit retains a full set of parameters and communicates with others in a ring-like logical topology (i.e., Figure~\ref{fig:intro:rar-arch}). The goal of both methods is to keep consistent parameter updates across all workers throughout the training phase while maintaining high throughput.

The popular PS architecture suffers from a critical bottleneck known as the ``incast'' issue~\cite{chen2023boosting}. This bottleneck emerges when numerous workers attempt simultaneous communication with a single PS, potentially leading to network congestion and consequently impeding the system's overall performance and scalability. The inherent limitations of PS have driven research towards alternative strategies that circumvent such challenges. In-Network Aggregation (INA) has been proposed as a solution to alleviate network congestion in PS. This approach, as incorporated into several works, including SwitchML~\cite{sapio2021scaling}, ATP~\cite{lao2021atp}, INAlloc~\cite{temp:inalloc}, and PANAMA~\cite{gebara2021network}, has significantly improved network congestion and DDL efficiency. However, the incremental deployment capabilities of PS-based INA approaches remain weak. Notably, PS-based INA methods necessitate the replacement of nearly all congestion-point switches in the topology for significant network throughput enhancements. For example, even if we replace 80\% of the regular switches with INA switches, the network throughput will be only 50\% (please refer to \S~\ref{sec:bom_model}).

The RAR architecture is another method to alleviate network congestion in the PS. RAR avoids network congestion issues due to its unique communication structure - a unidirectional loop. This design ensures that as long as the workers involved in the synchronization process are topologically linked, there will be no communication bottlenecks. However, a significant issue in RAR is the lengthy dependency chain, where the length of the chain is directly proportional to the number of participating workers. This dependency chain problem introduces a throughput degradation into the system as the chain extends. What is more, the performance degradation or downtime of even a single worker can significantly impact the overall system performance. For details, please refer to \S~\ref{sec:rar_dependency_chain}.

A novel architecture should be proposed to fully leverage the capability of in-network computing switches to alleviate traffic bottlenecks, thereby enhancing DDL training throughput while offering improved incremental deployment capability.
Our approach uses an agent-worker model to infuse \textbf{R}ing-AllReduce with \textbf{I}n-\textbf{N}etwork \textbf{A}ggregation \textbf{(\solution)} capabilities. The \solution\ design ingeniously amalgamates the INA switch's functionality into RAR's synchronization process, proposing a holistic architecture and workflow. \solution\ accommodates two types of workers - the abstracted worker (embodied by a \solution-enabled rack) and the autonomous worker (a regular RAR worker) - to ascertain both compatibility and peak throughput. Consequently, \solution\ achieves superior throughput while attaining a higher incremental deployment capability compared to PS-based INA methods, without being encumbered by RAR's long dependency chain issue. To the best of our knowledge, no existing work has yet incorporated INA capabilities into RAR. In this context, \solution\ emerges as a pioneer. 

In summary, this study makes the following contributions:

\begin{enumerate}
    % \item We introduce \solution, the first approach to integrate INA capability into RAR. This study represents a significant stride forward in the field of DDL training with data parallelism.
    \item We propose the Bandwidth-Occupation Model (BOM) to model existing PS-based INA methods. Through the BOM, we identify the lack of incremental deployment capabilities in PS-based INA methods. Concurrently, we analyze the problem of throughput degradation caused by the lengthy dependency chain in RAR.

    \item We introduce \solution, the first initiative to infuse RAR with INA capability. \solution\ leverages the agent-worker model to bring about the novel design, which uses the INA switch and the agent to abstract the workers under one rack. This results in \solution\ possessing superior incremental deployment capabilities compared to PS-based INA approaches. Additionally, the comprehensive design of \solution\ addresses the long dependency chain issue found in RAR by compressing the long dependency chain under a rack into a single hop, thus enhancing the throughput of RAR-based DDL. We implement a prototype of \solution\ on a P4 programmable switch.

    \item We conduct extensive evaluations on NS3 and a real testbed using five real-world DL models and four datasets. These evaluations affirm the effectiveness and efficiency of \solution\ in enhancing the performance of common DDL training tasks. Compared to the state-of-the-art INA approaches like ATP, \solution\ delivers a 50\% throughput advantage at an equivalent cost. Furthermore, in comparison with traditional RAR and PS, \solution\ can boost the throughput by up to 6x.
\end{enumerate}

% TODO
The remainder of this paper is organized as follows. \S~\ref{sec:background} reviews the background. \S~\ref{sec:motivation} presents the motivation and design concepts of \solution. \solution's design details are provided at \S~\ref{sec:design} and implementation details are provided at \S~\ref{sec:implementation}. \S~\ref{sec:evaluation} shares the evaluation results, followed by the discussion in \S~\ref{sec:discussion}. Related works are introduced at \S~\ref{sec:related_works}. We conclude the study in \S~\ref{sec:conclusion}.

% It is this problem that \solution aims to address, seeking to enhance RAR's robustness while retaining its innate advantage of avoiding network congestion.

% The development of deep learning, including artificial general intelligence (AGI), Large language generation models like GPT, and picture segmentation models like Segment Anything.

% Distributed deep learning is emerging with the increasing size of the dataset and model. Data parallelism are usually used for large dataset.

% PS and Ring-AllReduce are two commonly used parameter synchronization methods in data parallelism. Give a brief introduction to them.

% Since PS suffers from severe incast problem, INA are used in PS. List works like ATP, INAlloc, PANAMA.

% RAR has no congestion problem, for its communication is a one-way loop. This means that as long as the workers participating in the synchronization are topologically connected, there will be no communication bottlenecks. However, RAR suffers from another problem: a long dependency chain. The length of RAR's dependency chain is equal to the number of its workers, which is too long for poor robustness.

% To the best of our knowledge, there is no approach that introduces INA capability into RAR. Rina is the first to do this. Rina uses the agent-worker model to introduce INA capabilities into RAR. Rina can significantly alleviate the long dependency chain problem of RAR, while allowing incremental deployment to lower the equipment cost threshold for distributed training.

\section{Background}\label{sec:background}

\subsection{Distributed Deep Learning}

The mathematical goal of DL can be defined as the following optimization problem (Equation~\ref{eq:dl}), where $d_i$ is a data sample of dataset $D$, and $w$ represents all the parameters of the model and $y_i$ is the associated label with $d_i$. $f$ takes an input and outputs a prediction. $loss$ is the objective function. The goal is to minimize the average loss across the dataset.

\begin{equation}
% \vspace{-0.1in}
\begin{aligned}
& min\  \mathbb{E}_{d_i\in{D}}{loss(f(w, d_i), y_i)}
\end{aligned}
\label{eq:dl}
\end{equation}

With the rapid growth of the size of datasets and models, DDL has gained lots of research interest and has become the primary method to improve the training efficiency and throughput to meet the demands both industrially and academically.

There are two prominent parallelism schemes of DDL, data parallelism~\cite{shallue2018measuring} and model parallelism~\cite{forrest1987implementing,wang2022overlap}. Data parallelism duplicates training models across all computing workers. In a single iteration, each computing worker processes different mini-batches of data to calculate the local gradient updates which are exchanged with other workers later before updating the model parameters~\cite{le2011optimization}. When comes to data parallelism, Equation~\ref{eq:dl} changes into Equation~\ref{eq:ddl}, where $N$ denotes the number of workers. The synchronization in data parallelism is the main optimization objective of this study.

\begin{equation}
\begin{aligned}
& min\  \frac{1}{N}\sum_{i=1}^{N}\mathbb{E}_{d_i\in{D_i}}{loss(f(w, d_i), yi)}
\end{aligned}
\label{eq:ddl}
\end{equation}

Model parallelism splits the model parameters to multiple workers to make it possible to train large-size models. Each worker holds a subset of model parameters or layers. In every iteration, the sampled mini-batch of datasets is copied to all workers, and different parts of the DL model are computed on different workers. Model parallelism is also an important area of study, which is orthogonal to the data parallelism emphasized in this paper~\cite{shoeybi2019megatron}.

%But due to the DNN model's inner dependency between different layers, model parallelism has to overcome complicated computing subsequence thus leading to its difficulty to accelerate the training~\cite{mirhoseini2017device}.

\subsection{Synchronization Architectures}\label{subsec:synchronization_atchitectures}

\subsubsection{Parameter Server Architecture}

The PS architecture~\cite{li2014scaling} is a straightforward method for parallel computing across multiple workers (Figure~\ref{fig:intro:ps-arch}). In this architecture, a PS node maintains and manages a global model. During each training iteration, each worker computes its local gradients based on its own mini-batch and communicates these results to the PS. The PS updates the global model and synchronizes it with each worker. Typically, there are two synchronization models: Bulk Synchronous Parallel (BSP)~\cite{valiant1990bridging} and Asynchronous Parallelism (ASP)~\cite{lian2015asynchronous}. In BSP, workers must await a synchronization barrier before initiating the next iteration. Conversely, ASP removes this synchronous barrier. Generally, BSP tends to yield higher accuracy, while ASP significantly increases training throughput. Regardless of the synchronization model, the PS architecture remains a prevalent choice in large-scale training clusters.

\subsubsection{Ring-AllReduce Architecture}

\begin{figure}[tbp]
    \centering
    \subfigure[ScatterReduce stage.]{\includegraphics[width=0.48\columnwidth]{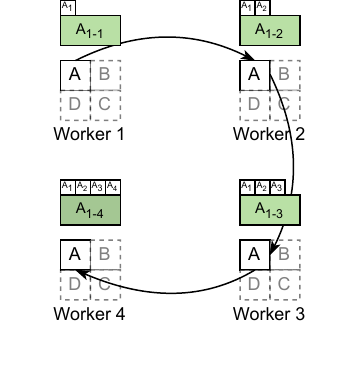}
    \label{fig:motivation:scatterreduce}}
    \subfigure[AllGather stage.]{\includegraphics[width=0.48\columnwidth]{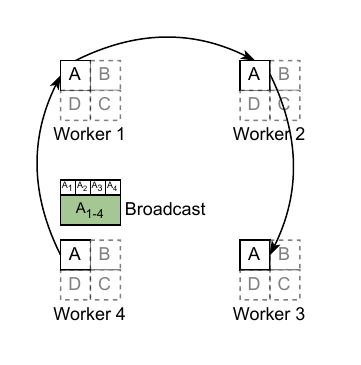}
    \label{fig:motivation:allgather}}
    \caption{Details for Ring-AllReduce.}
    \label{fig:enter-label}
    \vspace{-0.2in}
\end{figure}

AllReduce (AR) is a decentralized architecture proposed to alleviate communication bottlenecks in the PS architecture (Figure~\ref{fig:intro:rar-arch}). AR treats all machines as workers, thus eliminating the need for PS. Ring AllReduce (RAR) stands out among AR algorithms due to its superior bandwidth performance. RAR splits communication phases into ScatterReduce and AllGather. In the ScatterReduce phase, each of the $N$ workers divides their local gradients into $N$ chunks. Each worker, in every iteration, transmits a chunk to its neighbor, receives one, and adds it to the corresponding chunk. The chunks transmitted and received in each iteration are different, with each worker forwarding the chunk received in the previous iteration. After $N-1$ iterations, each worker possesses a globally updated chunk. For instance, in Figure~\ref{fig:motivation:scatterreduce}, worker 1 forwards chunk A to worker 2 in the first iteration. Worker 2 adds it to its local chunk and passes it to worker 3 in the subsequent iteration. After the ScatterReduce phase, worker 4 will possess a fully updated chunk A. During the AllGather phase, each worker transmits its complete chunk to the next worker and obtains one from the previous. Like in ScatterReduce, each worker forwards the chunk it received in the previous iteration. After $N-1$ iterations, every worker has fully updated gradients. As illustrated in Figure~\ref{fig:motivation:allgather}, worker 4 sends its fully updated chunk A to worker 1, who then forwards it to the next worker (i.e., worker 2). Thus, at the end of the AllGather phase, each worker possesses a fully updated result for all chunks. Unlike PS, RAR operates solely in BSP mode. While RAR achieves optimal bandwidth performance, it suffers from issues of extensive dependency chains and vulnerability to a single point of failure~\cite{mattson2020mlperf}.

\subsection{Bring INA into DDL}

In recent years, advancements in programmable networks have driven a surge of research employing INA techniques to expedite DDL training. INA leverages the computational power within programmable switches to aggregate gradients from multiple nodes, reducing network traffic and accelerating DDL training. Notable works in this domain include SwitchML~\cite{sapio2021scaling} and ATP~\cite{lao2021atp}, both of which aim to enhance overall training speed by offloading the gradient aggregation task to switches. In SwitchML, all gradients are aggregated within the switches, thus training speed hinges significantly on the switches' aggregation capabilities. ATP adopts a best-effort strategy for gradient aggregation, where gradients not aggregated at the switch level are relayed to the PS for aggregation. 

%Addressing the issue of low memory utilization efficiency in existing INA solutions, INAlloc~\cite{temp:inalloc} designs a memory management system that ensures full utilization of switch memory resources within a shared cluster, decreasing task deadline miss ratio and completion time.

Nevertheless, these optimization efforts are tailored specifically for INA within the PS architecture. The RAR architecture, renowned for its efficient communication performance, is gaining increased attention. To our knowledge, no existing research explores INA utilization within the RAR architecture, presenting a distinct set of challenges and opportunities at the core of this study.

\section{Motivation and Concepts of \solution}\label{sec:motivation}

In this section, we first propose to analyze the issues with RAR methods, specifically \textbf{their long dependency chains}. Next, we present the Bandwidth-Occupation Model (BOM) for all existing PS-based INA methods to illustrate their problem: \textbf{the lack of incremental deployment capability}. Finally, we present the design concepts and architecture of \solution, briefly elaborating on its advantages over both the state-of-the-art PS-based INA methods and traditional RAR methods.

\subsection{Long Dependency Chain Problem in Ring-AllReduce}\label{sec:rar_dependency_chain}

Compared to PS-based INA, RAR does not have communication bottlenecks, which are determined by the communication mode of RAR. The following provides quick proof.

\begin{enumerate}
    \item We can view a network as a connected undirected graph. Let $G = (V, E)$ be a connected undirected graph.
    \item Transform G into a directed graph $D = (V, D)$ by replacing each ${u, v} \in E$ with two directed edges $(u, v)$ and $(v, u)$ in $D$.
    \item By this transformation, for each vertex $v \in V$, the in-degree equals the out-degree.
    \item According to the Eulerian path and circuit conditions~\cite{bona2006walk} in directed graphs, since the in-degree equals the out-degree for all vertices in $D$, there exists an Eulerian circuit.
    \item Hence, a path in $D$ starts from any worker, visits every worker once, and finally ends at the starting worker. This guarantees the RAR's requirements for communication without bottleneck.
\end{enumerate}

\begin{figure}
    \centering
    \includegraphics[width=\columnwidth]{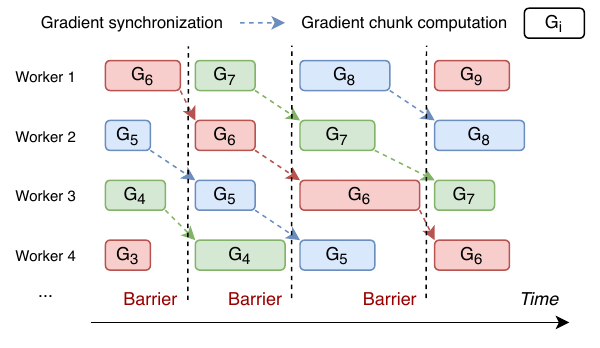}
    \caption{The node error will block the whole RAR synchronization process.}
    \vspace{-0.2in}
    \label{fig:rar_chain_challenge}
\end{figure}

Although RAR has been proven to be free of bandwidth bottleneck, it still suffers from the issue of long dependency chains~\cite{wan2020rat}. The most significant problem caused by long dependency chains is that throughput performance becomes affected by the increasing number of nodes. Take Figure~\ref{fig:rar_chain_challenge} as an example. The figure shows the workflow of an RAR pipeline. The same part of the gradients shares the same color. Each node sends its computed gradient $G_i$ to the next worker. However, according to the implementation of the latest MPI libraries such as NCCL~\cite{web:nccl} and OpenMPI~\cite{web:openmpi}, each round of communication has a barrier to global synchronization. Interference caused by load fluctuation, interrupts, garbage collection, or background tasks during Worker 3 processing $G_6$ will defer the global completion time of this step, Indicating that single-point failure will directly slow down the entire training process. This is the fundamental problem for the long dependency chain.

Suppose the whole cluster has $N$ workers. The system overhead of sending gradient (including network protocol, memory movement, et al.) chunk i is $O(G_i)$, while the computation and communication time is $C(G_i)$. For j-th round synchronization, the time consumption of worker $n$ will be $O(G_u) + C(G_u), u = (i + n)\%N $. Since the existence of barriers, the estimated time consumption of the ScatterReduce phase will be $\sum_{u=1}^N Max(O(G_u) + C(G_u))$. 

In practical scenarios, $O(G_u)$ can be considered a fixed overhead independent of $N$. Typically, the distribution of $C(G_u)$ is proportional to $N$ in a linear fashion. We assume $C(G_u)$ follows a normal distribution, where its mean is proportional to $N$ and the variance is a fixed value, expressed as $C(G_u) \sim \mathcal{N}\left(k \cdot \frac{1}{N}, \sigma^2\right) $. Here, $k$ is a constant. The standard deviation $\sigma_u$ is also assumed to be a constant $\sigma$. For a random variable $X$ that follows a normal distribution $\mathcal{N}(\mu, \sigma^2)$, the expected maximum value $M_n$ (taken from $n$ independent and identically distributed samples) can be approximated as $\mathbb{E}[M_n] \approx \mu + \sigma \sqrt{2 \ln n}$. Thus, the time consumption $T$ of RAR during the ScatterReduce phase can be expressed as:

\begin{equation} \label{eq:long_chain}
\begin{split}
T = & N \cdot O(G_u) + \sum_{u=1}^N Max(C(G_u)) \\
= & N \cdot O(G_u) + \sum_{u=1}^N \mathbb{E}[\max(C(G_u))] \\
\approx & N \cdot O(G_u) + k + N \cdot \sigma \sqrt{2 \ln N}
\end{split}
\end{equation}

From Equation~\ref{eq:long_chain}, it can be seen that the value of $T$ increases with the size of $N$, which demonstrates that the synchronization completion time of RAR increases as the number of nodes increases. It is also noteworthy that an increase in $\sigma$ will also lead to an increase in $T$, which means that if the nodes' performance is unstable, the synchronization completion time of RAR will be further prolonged. This phenomenon is common~\cite{li2014scaling,eisenman2022check, rojas2020study}, thus reducing the long dependency chain of RAR is highly necessary.

% However, if \textit{Worker 3} encounters an error while computing $G_6$, this single-point failure will directly slow down the entire training process. As \textit{Worker 3} experiences the error, \textit{Worker 4} will become idle after computing $G_5$, waiting for \textit{Worker 3} to send $G_6$. Similarly, \textit{Worker 2} will face a bottleneck as $G_8$ piles up at \textit{Worker 3} or fails to transmit due to the unfinished $G_6$. This will severely impede the entire training process. More importantly, as the number of nodes increases, the number of steps RAR needs to perform also increases, leading to a higher probability of single-point failures.

% It is worth noting that node errors occur very frequently during the DDL process~\cite{li2014scaling}. The probability of node errors increases with the length of the training time and the number of nodes involved in the training~\cite{eisenman2022check, rojas2020study}. In addition to issues with the nodes' own computational resources, network bandwidth fluctuations or abnormal data packet loss can also lead to consequences similar to node errors. This means that the issue of RAR's dependency chain length cannot be ignored.

As a widely used AllReduce method in the industry, Hybrid Allreduce (H-AR)~\cite{jia2018highly} has been proposed to address the issue of long dependency chains through a multi-step AllReduce process. H-AR first performs a ScatterReduce within the ToR, then an AllReduce between ToRs, and finally an AllGather within the ToR. This approach indeed mitigates the long dependency chain problem and achieves better performance than RAR. However, Rina's utilization of INA switches can achieve even higher throughput compared to H-AR, for \solution\ not only mitigates the dependency chain but also provides in-network computation capabilities. A detailed comparison is provided in \S~\ref{sec:evaluation_on_throughput}.

\subsection{Modeling PS-based INA with BOM}\label{sec:bom_model}

A major advantage of the PS-based INA approaches in improving the throughput of DDL training tasks is its ability to reduce network traffic~\cite{gebara2021network}. In this section, we quantify this through the BOM model and discuss their weakness in incremental implementation.

\textbf{Assumptions:} The DDL training cluster uses the BSP synchronization algorithm. All nodes need to send the gradients of their local models to the PS synchronously, followed by a broadcast generated by the PS. The INA switch can fully aggregate incoming traffic (as proven to be feasible in INAlloc~\cite{temp:inalloc} under the single-job scenario). If the corresponding PS-based INA method does not require a PS server, the farthest INA switch is treated as the PS. The entire topology is homogeneous, with a link bandwidth of $B_0$. There is no multipath scenario in the topology, that is, there is exactly one path from all nodes to the PS.

\textbf{Lemma 1:} \textit{For a topology only containing regular switches and $n$ workers, the worker throughput is $1/n$.}

\begin{figure}[tp]
    \centering
    \includegraphics[width=\columnwidth]{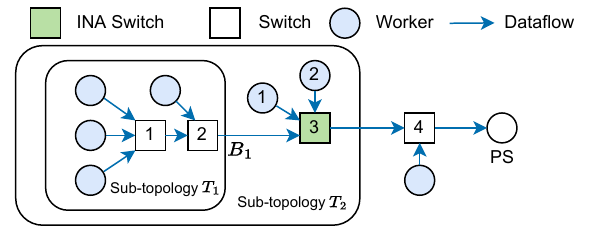}
    \caption{A sample of traffic on the switch under PS architecture.}
    \label{fig:bom_model}
    \vspace{-0.2in}
\end{figure}

As shown in the sub-topology $T_1$ in Figure~\ref{fig:bom_model}, this topology does not include any INA switches. Assuming the throughput of its outbound switch 2 is $B_1$, the worker throughput in this topology is $B_1/4$. The proof is as follows.

Assume a complex topology $T$. From the topology $T$, we select the PS node working as the root to build a Directed Acyclic Graph (DAG) tree, which can be used to represent the network traffic during the gradient aggregation phase. 

Given a DAG tree $G = (V, E)$, where $V$ is the set of vertices (or nodes) and $E$ is the set of directed edges. $G$ is a subtree of $T$. The root node is denoted as $r$ and $L$ is the set of leaf nodes. The output rate of a node $v$, denoted as $OR(v)$, is defined as the number of outgoing edges from $v$.

% For all leaf nodes $l$, we have $OR(l) = 0, \forall l \in L$s.

We aim to prove that $\forall l \in L$, $OR(l)$ is determined by the output rate of the root node $r$, $OR(r)$. To do this, we use the principle of mathematical induction.

\emph{Base case:} When $|V| = 1$ (i.e., the tree only contains the root), it's trivial that $OR(l)$ depends on $OR(r)$ since they are identical.

\emph{Additional case:} We select a non-leaf node $v$ in \textit{inductive step}. When $|V| = 2$, no leaf nodes exist, thus only one of the leaf nodes can be selected arbitrarily. In this scenario, $OR(l) = OR(r)$ is also evident.

\emph{Inductive step:} Assume the proposition holds for any tree with $|V| = n$, i.e., for any tree with $n$ nodes, $\forall l \in L$, $OR(l)$ is determined by $OR(r)$.

We need to prove that for any tree with $|V| = n + 1$, $\forall l \in L$, $OR(l)$ still depends on $OR(r)$.

Consider a tree $G$ with $|V| = n + 1$. Select a non-leaf node $v$ (except $r$) and consider the sub-tree $G'$ formed by removing one of $v$'s child nodes $c$ (and edges attached to $c$). Now $G'$ has $n$ nodes.

By the induction hypothesis, $\forall l' \in L'$ (leaf nodes in $G'$), $OR(l')$ depends on $OR(r)$. Since removing the child $c$ of $v$ doesn't change $OR(r)$, it still holds that $\forall l' \in L'$, $OR(l')$ depends on $OR(r)$.

For the removed node $c$, since it was an outgoing edge from $v$ and eventually from $r$, $OR(c)$ also depends on $OR(r)$.

Hence, we have shown that for any tree $G$ with $|V| = n + 1$, $\forall l \in L$, $OR(l)$ is determined by $OR(r)$. We can conclude that for any topology only containing regular switches, the output rate of each worker is determined by the number of outgoing bandwidth from the root, which is $OR(r)/W$. $W$ represents the number of workers

\textbf{Lemma 2:} \textit{The INA switch and its children can be viewed as one worker.}

INA switches can aggregate the received gradients and output dataflow without redundant positional gradients. In this way, to the parent node of this INA switch (no matter the other INA switch or PS), its behavior is manifested as an independent worker.

\textbf{Lemma 3:} \textit{For an INA switch, the actual throughput depends on the worst-performing child.}

Take Figure~\ref{fig:bom_model} as an example. For the outbound INA switch 3 in topology $T_2$, even though it has sufficient INA capabilities to allow workers 1 and 2 to function at 100\% throughput, it is still limited by the slowest child node (topology $T_1$). This implies that the actual outbound bandwidth of topology $T_2$ is $B_0/4$, which is also the actual global throughput of this example.

At this point, given the topology, nodes, and placement of INA switches, we can calculate the actual throughput of all the workers in this cluster during the synchronization phase, based on BOM. Take Figure~\ref{fig:bom_model} as an example. For switch 4, as its child nodes consist of an INA switch (treated as one worker) and a worker, the limitation on the throughput imposed by switch 4 is $B_0/2$. For switch 3, since it is an INA switch, workers 1, 2, and switch 2 can all send gradients to it at 100\% throughput. This implies that $B_1=B_0$. For sub-topology $T_1$, based on Lemma 1, the actual throughput limit is $B_0/4$. The actual throughput of all workers globally is taken as the minimum value, which is $B_0/4$.

\subsection{Incremental Deployment is Challenging for PS-based INA}

\begin{figure}[tbp]
    \centering
    \subfigure[Fat-tree (k=4).]{\includegraphics[width=0.36\columnwidth]{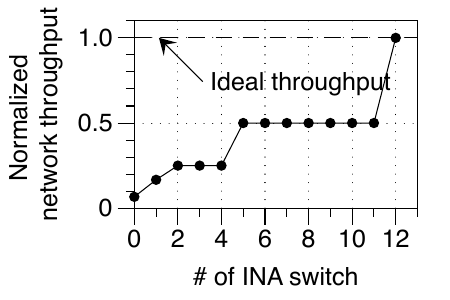}
    \label{fig:motivation:li_fat-tree}}
    \subfigure[Dragonfly (a=4, g=9, and h=2).]{\includegraphics[width=0.6\columnwidth]{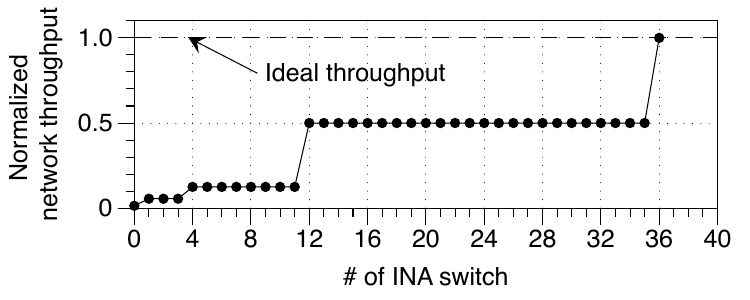}
    \label{fig:motivation:li_dragonfly}}
    \caption{Lack of incremental deployment capability for PS-based INA.}
    \label{fig:lack-of-inc}
    \vspace{-0.2in}
\end{figure}

The incremental deployment capability of PS-based INA methods is poor. Using BOM, we evaluate the changes in DDL training throughput in a specific topology scenario, starting from ``all switches are regular switches'' to replacing all switches in the topology with INA switches.

The training task we selected is ResNet50~\cite{resnet-he2016deep}, using the CIFAR-10~\cite{cifar10-krizhevsky2009learning} dataset. We have chosen two topologies, namely the standard Fat-tree~\cite{al2008scalable} (k=4) and standard Dragonfly~\cite{kim2008technology} (a=4, g=9, and h=2) topologies, which are commonly used in data centers. The corresponding results are shown in Figure~\ref{fig:lack-of-inc}. As can be seen, to achieve significant throughput improvements, PS-based INA methods need to replace regular switches in the entire network with high-performance programmable switches as much as possible. If only a part of the switches is replaced, the effect of INA cannot be well-utilized. Therefore, designing an incremental deployment-friendly DDL synchronization architecture is one of the important design principles of \solution.

% As a result, a major goal of \solution\ is to reduce RAR's dependency chain to prevent overall throughput from decreasing due to node errors. We also design a protocol to mitigate the throughput degradation caused by sporadic node errors.

% However, RAR is not without its flaws. The most serious problem with RAR is its long dependency chain, which is positively correlated with the number of nodes~\cite{wan2020rat}. The straggler problems that occur in the DDL training process are not rare~\cite{li2014scaling}. A longer dependency chain can amplify these errors that may occur during DDL training, thereby reducing the throughput and robustness of RAR training. Therefore, \solution\ tries to incorporate the INA device into RAR. One of the significant design goals of \solution\ is to introduce INA to improve the throughput and reliability of RAR-based training.

\subsection{\solution\ Design Concepts and Challenges}

Before introducing details of \solution, we summarize the existing PS-based INA methods from a higher perspective. The existing PS-based INA involves the INA device wrapping all its attached devices into a unified external device. From the viewpoint of other devices after the INA device, the INA and its workers are combined as a larger-scale worker.

If we aspire to incorporate the same INA capability into RAR and resolve the issues of long dependency chains, we should adopt a method similar to PS-based INA. \textbf{This involves considering the INA switch and its attached nodes as a whole, that is, as an abstracted worker.} The design of \solution\ follows this concept, and here are the other challenges.

\begin{enumerate}
    \item The workflow of RAR is significantly more complex than PS, making the integration of INA into RAR a challenging task. How should the architecture and workflow of \solution\ be designed?
    \item The introduction of INA switches will lead to new design issues, such as congestion control and reliability assurance. How does \solution\ address these issues?
    \item The INA method based on PS lacks incremental deployment capability. How can \solution\ achieve incremental deployment capability? How should switches be incrementally deployed?
\end{enumerate}

In the following section, we will provide a detailed introduction to the design of \solution\ addressing these challenges.

\section{Design Details of \solution}\label{sec:design}

\solution\ is designed to incorporate the INA switch into DDL training tasks using the RAR synchronous architecture as a basis. \solution\ not only retains the benefits of the RAR architecture, including the absence of communication bottlenecks, but it also effectively mitigates the issue of the long dependency chain. Essentially, \solution\ upholds the RAR workflow pattern, which includes the ScatterReduce and AllGather phases, and optimizes these stages to leverage the capabilities of INA devices. Moreover, \solution\ introduces a new workflow based on the agent-worker model, discussed in \S~\ref{subsec:agent-worker-model}. This model allows INA devices to manage all workers within each rack, as detailed in \S~\ref{sec:rina_dataflow}. We implement a lightweight congestion control protocol to meet the unique requirements of \solution\ and its utilization of INA capabilities. Furthermore, when compared to PS-based INA, \solution\ provides superior incremental deployment capabilities. We discuss in detail in \S~\ref{subsec:incremental_deployment} how \solution\ achieves incremental deployment capability and how incremental deployment is carried out.

\subsection{Agent-worker Model}\label{subsec:agent-worker-model}

\begin{figure}[tbp]
    \centering
    \includegraphics[width=\columnwidth]{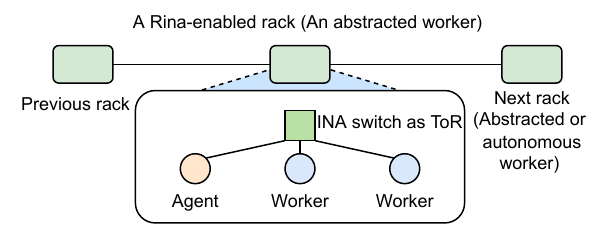}
    \caption{Agent-worker model for \solution.}
    \label{fig:design:agent-worker-model}
    \vspace{-0.2in}
\end{figure}

Figure~\ref{fig:design:agent-worker-model} presents a comprehensive illustration of the agent-worker model. \textbf{\solution\ adopts the rack as its operative unit (i.e., an abstracted worker), superseding each rack's ToR switch with an INA switch to enable the INA capability.} When viewed from an individual rack's perspective, once its ToR switch is supplanted by an INA switch, the worker of the lowest rank within that rack assumes the responsibility of managing the rack's INA switch and its workers. This low-rank worker is termed the ``agent''. In addition to performing computational tasks, the agent also performs several additional functions, such as initiating \solution, assigning tasks to the INA switch, and provocation of synchronizations for other workers. For real-world implementations, these agent roles are executed via an extra daemon program that runs on one of the workers in the group (usually the first worker).

\begin{figure*}[tbp]
    \centering
    \includegraphics[width=1.9\columnwidth]{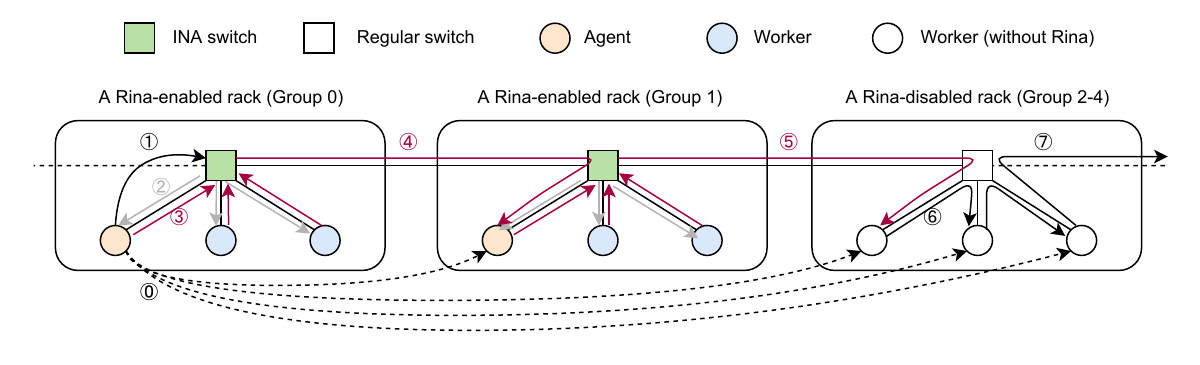}
    \vspace{-0.2in}
    \caption{Architecture and dataflow of \solution's ScatterReduce phase.}
    \label{fig:design:rina-arch}
    \vspace{-0.2in}
\end{figure*}

\subsection{\solution's Architecture, Workflow, and Dataflow} \label{sec:rina_dataflow}

As illustrated in Figure~\ref{fig:design:rina-arch}, \solution\ is an architecture facilitating the harmonious operation of regular and INA switches. If a rack with an INA-enabled ToR switch exceeds two nodes, \solution\ can be deployed, designating such a rack as an abstracted worker and \solution-enabled rack. Rather than assigning tasks individually, \solution\ adopts a group-based approach, considering \solution-enabled racks as an abstracted worker. Conversely, for \solution-disabled racks, each worker is regarded as an autonomous worker or an autonomous group.

\subsubsection{Model and Dataset Partitioning} 

Data-parallel DDL training usually necessitates dataset partitioning to attain parallelism, with the addition of model splitting by RAR to guarantee synchronization throughput. Consequently, \solution\ calls for meticulous consideration in the partitioning of data and models. In both PS and RAR architectures, when all workers share equivalent computational capacity, an equal dataset segment is allocated to each worker to approximate global computation time. Conversely, with workers possessing heterogeneous computational capabilities, strategies such as batch-size tuning~\cite{chen2019round, chen2023osp} are employed to synchronize computation time.

Regarding model partitioning during synchronization, the conventional RAR strategy divides the model evenly across workers to ensure near-equal synchronization time per step (refer to \S~\ref{subsec:synchronization_atchitectures}). However, \solution\ necessitates model partitioning in line with the number of groups. This stipulation arises primarily because each \solution-enabled rack, represented as an independent worker by its respective agent and INA switch, must be accounted for.

At the onset of training, both the dataset and model partitions are disseminated to ensure uniform initial parameter states across all workers. We designate a global control node, which will be the agent of the 0th group or 0th autonomous worker. As training commences, this node randomizes all parameters of the target model and transmits the partition data of both the dataset and model to all groups (Figure~\ref{fig:design:rina-arch}-(\textcircled{\small 0}). To expedite the distribution of control messages within a \solution-enabled rack, \solution\ utilizes multicast. Once this preparatory phase is complete, all workers embark on their training tasks.

\subsubsection{Synchronization Process} 

Before each round of DDL training, synchronization is necessitated for all workers to synchronize training results based on each node's respective dataset. This periodic process aligns well with near-equal computation times across nodes, enabling all workers to commence synchronization approximately simultaneously. As in the RAR system, upon completion of computation, each node promptly transmits its model partition's gradient to the subsequent worker while concurrently receiving the gradient from the prior worker and conducting local aggregation. This attribute is retained in \solution. For autonomous workers, their actions mirror those in RAR. Meanwhile, within \solution-enabled racks managed by the agent, the agent handles task delegation.

Similar to RAR, \solution\ is divided into two phases: ScatterReduce and AllGather. During the ScatterReduce phase, all nodes undergo a synchronization round, ensuring each node acquires the final computation result for a model gradient segment. This gradient portion is then broadcasted to all nodes in the AllGather phase. Participation of the INA switch in aggregation is necessary for ScatterReduce, while AllGather necessitates switch support in multicast.

In a cluster comprising $N$ nodes, these two phases would necessitate $2(N-1)$ steps in the RAR system. In contrast, in \solution\ with $G$ groups, each phase demands $2G-1$ steps. Given that a group may include several workers, in a cluster where each rack hosts eight computing nodes, $G$ would equate to $N/8$. Thus, the synchronization steps demanded by \solution\ are notably fewer than those required in RAR, which contributes to a higher throughput.

\subsubsection{ScatterReduce Phase} 

Given the inability of prevalent P4 programmable switches such as Tofino-1~\cite{web:tofino} to autonomously generate packets, and considering synchronization requirements, the synchronization signals of workers are uniformly triggered by the agent for each group.

Referring to Figure~\ref{fig:design:rina-arch}, upon the agent's completion of its current computation round, it relays aggregation task data to the switch (\textcircled{\small 1}). This information comprises the model range and size destined for the next group, the reserved memory space in the switch, and the ID of the node currently engaged in the aggregation. On receipt of this data, the switch converts this packet into a data pull message (\textcircled{\small 2}), which is multicast to all workers (including the agent) under this rack. Triggered by the pull message, all nodes initiate the transmission of corresponding gradients to the ToR switch (\textcircled{\small 3}). These parameters are then aggregated at the ToR switch and dispatched to the agent of the subsequent group (\textcircled{\small 4} and \textcircled{\small 5}). When the next group's agent receives the aggregated results from (\textcircled{\small 4} and \textcircled{\small 5}), it combines them with the corresponding local gradients. These gradient portions are then used for synchronization in the ensuing stage.

For autonomous workers, the synchronization mechanism of \solution\ necessitates only minor adjustments to RAR. As illustrated in Figure~\ref{fig:design:rina-arch}, when the subsequent hop of an abstracted worker is an autonomous worker, this part of communication devolves to a standard RAR operation (\textcircled{\small 6}). Conversely, while transmitting data from an autonomous worker to the abstracted worker, the workflow stays the same as the synchronization between abstracted workers. This guarantees minimal alterations to existing RAR within \solution\ and supports compatibility with regular RAR.

\begin{figure}[tbp]
    \centering
    \includegraphics[width=\columnwidth]{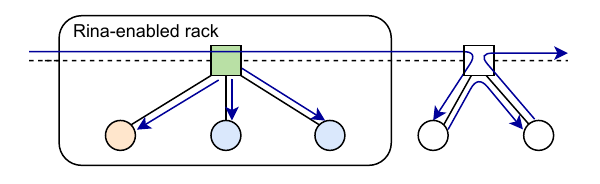}
    \caption{Dataflow of \solution's AllGather phase (Legends are the same as Fig.~7).}
    \label{fig:design:rina-allgather}
    \vspace{-0.2in}
\end{figure}

\subsubsection{AllGather Phase} 

When a specific partition of the gradients has been globally aggregated (i.e., after traversal through $G$ groups), it promptly enters the AllGather phase. During this stage, the gradients are broadcasted to all workers via a ring propagation path. In \solution, we harness the capabilities of INA switches to facilitate multicast, thereby enhancing the AllGather phase's performance.

As depicted in Figure~\ref{fig:design:rina-allgather}, when an agent obtains a gradient shard with the same ID as the one it initially synchronized with, the AllGather process for that gradient commences. This agent then forwards the gradient shard to the next group's agent. If the subsequent group is an autonomous worker, it will locally store the gradient shard and further pass it on to the next worker, which is the same as RAR. However, if the next group is an abstract worker, the corresponding INA switch will multicast this gradient to all its workers and the subsequent group. This design allows for faster broadcasting of the aggregated gradients to all workers compared to RAR.

\begin{figure}[tbp]
    \centering
    \includegraphics[width=0.9\columnwidth]{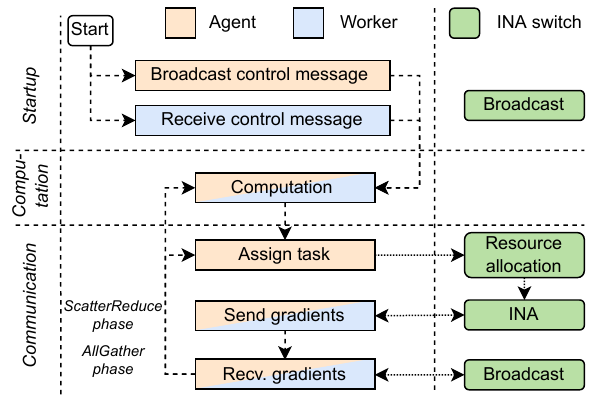}
    \caption{The workflow of \solution.}
    \label{fig:design:rina-flowchart}
    \vspace{-0.2in}
\end{figure}

The complete workflow, which encompasses task initiation, computation, and the synchronization process, can be referred to in Figure~\ref{fig:design:rina-flowchart}.

\subsection{Congestion Control and Reliability} \label{subsec:cc_reliability}

\subsubsection{Congestion Control}

The implementation of effective congestion control can mitigate network congestion and control the memory bottleneck of INA switches. Our \solution\ designs distinct congestion control mechanisms for ScatterReduce and AllGather. Given that the INA switch and the node maintain a one-hop distance in \solution, the congestion control scheme can remain straightforward. Nonetheless, it necessitates cross-rack congestion prevention measures to avert substantial packet loss.

As illustrated in Figure~\ref{fig:design:rina-arch}, congestion control initiates when (\textcircled{\small 2}) prompts workers in the \solution-enabled rack to deliver gradient shards to the INA switch. During the ScatterReduce phase, workers transmit at full speed~\footnote{``Full speed'' here denotes the speed aligned with the INA's capability. For instance, for the Tofino-1 P4 switch with 100G ports, due to hardware restrictions, the INA switch typically achieves a speed of around 20Gbps~\cite{he2023generic}.}, paralleling DCQCN's congestion control~\cite{zhu2015congestion}. Once the aggregate message reaches the following rack's agent, it returns an ACK. At this stage, congestion control employs an Additive Increase/Multiplicative Decrease (AIMD) approach, ceasing window expansion upon reaching ``full speed''.

During the AllGather phase, as other gradient shards' ScatterReduce phase might not have concluded, step (\textcircled{\small 4}) in Figure~\ref{fig:design:rina-arch} monopolizes the agent's downstream bandwidth. Consequently, the sending rate at this juncture starts from zero. The ACK is issued either by the autonomous worker or the agent of the abstracted worker.

\subsubsection{Reliability}

Despite recent methods' inability to manage worker errors, the PS-based INA method's central management node offers a significant advantage. It allows for active node exclusion upon error detection, enhancing overall reliability. Such a feature is lacking in the RAR approach due to its Peer-to-Peer (P2P) architecture. The P2P nature of RAR results in difficulties in making node replacement upon error detection challenging. However, the architecture of \solution\ is designed to effectively manage errors. We classify node errors into two categories: agent errors and worker errors. Upon the occurrence of an agent error, the other workers in the corresponding rack can default to regular RAR, ensuring uninterrupted training. If a worker error occurs and the worker is part of a \solution-enabled rack, its corresponding agent can promptly detect and exclude the faulty node from subsequent aggregation processes. If the worker is autonomous, other workers will automatically bypass the node. 

%We disseminate complete topology information to all workers at the start of synchronization, improving overall system reliability.

% Although the latest methods do not handle worker errors, the PS-based INA method has a central management node, making it relatively easy to actively exclude nodes when errors occur to reach higher reliability. This is almost impossible for RAR, which is mainly due to 1) RAR is a Peer-to-Peer (P2P) architecture, lacking a central management node that can discover and handle errors in real time; 2) RAR's workers basically only communicate with neighbors, which leads to the fact that the neighbor nodes do not know who to find as a replacement once the error occurs.

% The architecture of \solution\ ensures that it can handle errors well. In \solution, node errors are mainly divided into two categories, namely agent errors and worker errors. When an agent error occurs, the other workers in the corresponding rack can degrade to regular RAR to ensure that the training proceeds normally. When a worker error occurs, if the worker belongs to a \solution-enabled rack, its corresponding agent can quickly discover and exclude the node in the subsequent aggregation process. If the worker is an independent group, other workers will actively skip the node. We will pass the information of the entire topology to all workers when starting to synchronize global information, to ensure better reliability.

\subsection{Incremental Deployment} \label{subsec:incremental_deployment}

\solution's agent-worker model can integrate all nodes under an in-network computing switch into a single worker. This means that replacing a conventional ToR switch connecting $N$ nodes with an INA switch can reduce the length of the RAR dependency chain by $N$. Therefore, in the initial deployment phase, we should prioritize replacing normal switches with the most connected workers with INA switches.

As deployment progresses, we should consider replacing other normal switches in the topology that are not ToR switches. Refer to \S~\ref{sec:bom_model}, constructing a minimum spanning tree rooted at a specific worker node that connects all worker nodes. Then, by treating an INA switch and all its downstream worker nodes as a single worker, we can similarly replace the conventional switch with the most downstream nodes with an INA switch.

Through the gradual replacement of standard switches with those developed using our method, we can achieve effective incremental deployment. Each instance of switch replacement leads to a noticeable enhancement in the throughput of DDL tasks. A thorough evaluation of our incremental deployment capabilities is provided in Section \ref{sec:evaluation_incremental_capability}.

% The \solution\ boasts great incremental deployment capabilities, fundamentally ingrained in its design philosophy and architecture. In contrast to PS-based INA, which necessitates the costly and sensitive procedure of replacing all switches with INA switches, \solution\ implements a more cost-effective strategy. It accommodates the co-existence of conventional RAR nodes and \solution-enabled racks, highlighting its high compatibility with RAR. This approach ensures that each INA switch replacement in \solution\ diminishes the dependency chain of RAR, a measure with substantial benefits. We provide a comprehensive evaluation of incremental deployment capabilities in \S~\ref{sec:evaluation_incremental_capability}.

% \subsection{Utilize the Reverse Link After ScatterReduce}

% \begin{figure}[tbp]
%     \centering
%     \includegraphics[width=\columnwidth]{Figures/Design/Rina_Arch-Rina-AllGather.pdf}
%     \caption{\solution\ can utilize the reverse link during AllGather.}
%     \label{fig:design:rina-allgather}
% \end{figure}

% \section{Mathematical Analysis of Different Synchronization Architectures}

% \subsection{Time Consumption Analysis for PS and RAR}

% \subsection{INA Device Modeling}

% \subsection{Analysis for PS-based INA Approaches}

% \subsection{Analysis for \solution}

\section{Implementation}\label{sec:implementation}

We implement the \solution\ prototype on both P4 programmable switches~\cite{web:tofino} and workers. The deployment on the switches emphasizes INA capabilities, whereas the implementation on worker nodes principally aims to enhance the processing performance of small packets.

\subsubsection{P4 Programmable Switch}

Given the absence of floating-point computation capabilities in P4 switches, \solution\ adopts a method analogous to ATP~\cite{lao2021atp}, where all floating-point numbers are multiplied by an integer and then converted to integers at the worker nodes. This conversion enables the P4 switch to transform floating-point addition into a simpler integer addition operation. Moreover, considering hash-based memory allocation algorithm may cause collisions~\cite{lao2021atp}, \solution\ allocates contiguous memory space directly for INA tasks. To augment the INA throughput, we utilize the P4 switch's recirculate feature to extend the length of each INA packet.

% In the implementation of P4 programmable switches, we thoroughly consider their robustness and error-handling capabilities. The P4 switch has a simple MAC-based forwarding program in the first stage. This program allows the P4 switch to normally process other data packets and ensures that the switch can degrade to a regular switch when errors might occur in \solution\ (although in our evaluation, errors did not occur). The following stages are used to solve the INA logic. We also use the resubmit feature of the P4 switch to increase the length of each INA packet to increase the INA throughput.

\subsubsection{Worker}

We implement a \solution\ prototype as middleware, seamlessly integrated into PyTorch. By utilizing UDP, we develop a customized transport protocol that leverages Mellanox (NVIDIA) Raw Ethernet Programming~\cite{web:raw-ethernet-programming} to expedite the processing of user-space data packets. To enhance packet processing performance, we also employ TCP segmentation offload (TSO)~\cite{web:mellanox-TSO}.

% \subsubsection{Minor Implementation Details}

\section{Evaluation}\label{sec:evaluation}

\input{Contents/EvaluationFigures}

In this section, we evaluate the advantages of \solution\ compared to commonly used DDL training synchronization architectures, through simulation experiments and testbed experiments. The evaluation includes throughput, incremental deployment capabilities, and robustness. 

% Since there is no data loss in the synchronization architecture, \solution\ will not harm the final accuracy of the DDL training tasks.

\subsection{Evaluation Setup}

\subsubsection{Simulator}

We use the popular NS3~\cite{web:ns3} simulator as our simulation tool. We developed real worker nodes and PS in NS3 and implemented the logic of PS and RAR synchronization architectures. For switches, we use the switch component to simulate regular switches and use nodes to implement the simulation logic of INA switches. In the simulation, we additionally evaluate the incremental deployment capabilities of various methods in popular data center topologies. These topologies include standard Fat-tree~\cite{al2008scalable} (k=4) and standard Dragonfly~\cite{kim2008technology} (a=4, g=9, and h=2).

\subsubsection{Testbed Configuration}

We evaluate \solution\ using an 8-node cluster. The nodes are separated into 2 racks, for each rack has 4 nodes. They are interconnected through two Intel Tofino-1 P4 programmable switches (with 32x100Gbps ports) as the ToR switch. Each node has one AMD Epyc 7643 CPU (48 cores, 96 threads), 128GB RAM, and one Mellanox ConnectX-6 Ethernet Network Adapter with 2 100Gb ports. Additionally, each node has one NVIDIA RTX3090 GPU. The NVIDIA driver version is 460.91.03, and the CUDA~\cite{cuda} version is 11.2. The operating system is Ubuntu 20.04.2 with kernel version 5.15.0-75-generic. We use these 2 switches and 8 workers to build a spine-leaf simple topology for the evaluation of \solution.

\subsubsection{Workload}

In the evaluation, we conduct several experiments to evaluate the performance of 5 DL models and 4 datasets. The workloads include training ResNet50~\cite{resnet-he2016deep} and VGG16~\cite{vgg-simonyan2014very} models on the CIFAR10~\cite{cifar10-krizhevsky2009learning}, InceptionV3~\cite{inceptionv3-xia2017inception} on the CIFAR100~\cite{cifar10-krizhevsky2009learning}, ResNet101~\cite{resnet-he2016deep} on the ImageNet1K~\cite{deng2009imagenet}, and the BERTbase~\cite{devlin2018bert} model on the SQUAD1.1~\cite{rajpurkar2016squad}. The task for the BERTbase model is the fine-tuning task on the SQUAD1.1 dataset. The batch sizes for all image classification models are set as 64, while BERT is 12. In all results, the throughput unit for BERTbase is one question-answer pair per 10 seconds. All other hyper-parameters stay as default. These workloads are chosen to enable a thorough understanding of the strengths and limitations of \solution.

\subsubsection{Baseline and Metric}

We compare \solution\ with regular PS, RAR, H-AR~\cite{jia2018highly}, and PS-based INA (ATP~\cite{lao2021atp}). The PS-based approaches use co-located PS for better performance. H-AR is a widely used method in the industry, achieving improved parallel performance than RAR. The two INA methods ATP and \solution\ ensure that the INA switches used have no memory bottlenecks and have similar aggregation throughput. We do not present the evaluation of SwitchML~\cite{sapio2021scaling} since its throughput is consistently inferior to ATP. We primarily compare the performance differences in throughput of these four schemes, as well as the incremental deployment capabilities of ATP and \solution\ under different topologies. Specifically, we evaluate their gains in throughput by gradually replacing the regular switches in the corresponding topology with INA switches.

\subsection{Evaluation on Throughput}\label{sec:evaluation_on_throughput}

We assess \solution's throughput using five distinct DL models across both Fat-tree and Dragonfly topologies. Initially, we establish PS and RAR as baselines and juxtapose ATP and \solution\ at 50\% and 100\% INA switch replacement rates. In the Fat-tree topology, 50\% replacement entails using 6 INA switches in ATP and 4 in \solution, while for Dragonfly, it implies employing 18 INA switches in both methods. As depicted in Figure~\ref{fig:evaluation:throughput}, \solution\ significantly exceeds the common baselines of PS and RAR. 
H-AR outperforms RAR in terms of performance, but \solution\ can achieve better throughput than H-AR by replacing only half of the network switches. Compared to ATP, \solution\ can significantly enhance throughput by replacing 50\% of the network switches with INA switches. Furthermore, after replacing all switches with in-network computing switches, \solution\ performs comparably to ATP and even surpasses ATP in the ResNet50 model. This indicates that the integration of INA switches in \solution\ yields superior benefits and elevates overall performance. For ease of presentation, we express the throughput of the BERTbase model as Questions and Answers (QAs) every 5 seconds.

\begin{figure}[tbp]
    \centering
    \subfigure[Fat-tree (k=4).]{\includegraphics[width=0.95\columnwidth]{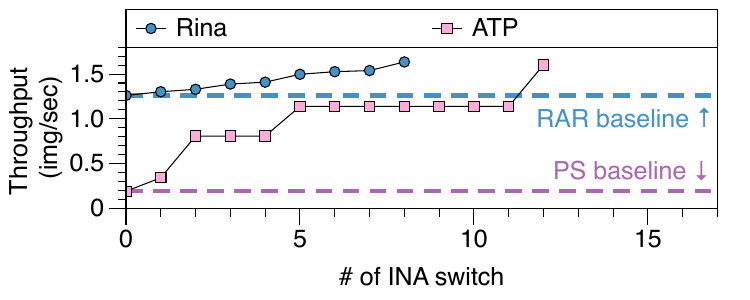}
    \label{fig:incre:fattree}}
    \subfigure[Dragonfly (a=4, g=9, and h=2).]{\includegraphics[width=0.95\columnwidth]{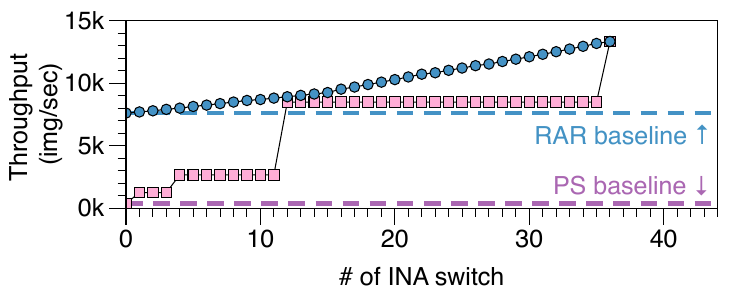}
    \label{fig:incre:df}}
    \caption{Evaluation on incremental capability.}
    \label{fig:evaluation:incre}
    \vspace{-0.2in}
\end{figure}

\begin{figure}[tbp]
    \centering
    \includegraphics[width=0.92\columnwidth]{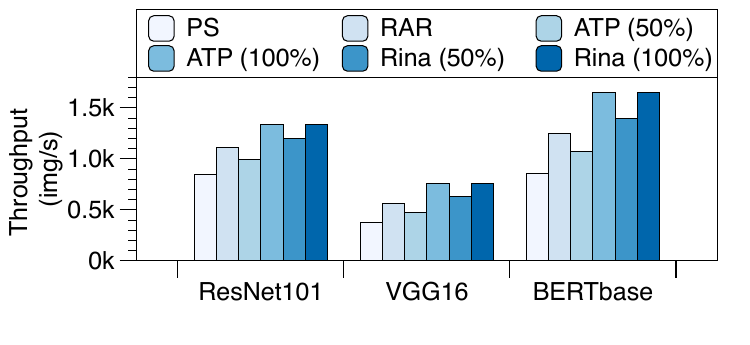}
    \caption{Testbed verification with 8 workers.}
    \vspace{-0.2in}
    \label{fig:thru:testbed}
\end{figure}

\subsection{Evaluation on Incremental Capability} \label{sec:evaluation_incremental_capability}

We further evaluate the incremental deployment capabilities of ATP and \solution\ using ResNet50, featuring a model with 98MB parameters. In both topologies, we progressively replace all switches with INA switches and measure their throughput. As illustrated in Figure~\ref{fig:evaluation:incre}, the throughput of \solution\ gradually increases as the count of INA switches rises. On the other hand, ATP, due to its lack of incremental deployment capabilities, only witnesses a throughput boost after a significant quantity of INA switch replacements. This indicates that with \solution, DDL training operators can experience performance enhancements proportional to their investment, thus offering substantial hardware cost-effectiveness.

\subsection{Testbed Verification}

We assess \solution's performance on a testbed, with the outcomes depicted in Figure~\ref{fig:thru:testbed}. The results reveal that \solution\ mirrors the performance advantages observed in the simulation. Its performance remains closely matched with ATP when all switches are replaced by INA switches. However, when only a fraction of switches are replaced, \solution\ can attain a training throughput surpassing ATP. This suggests that \solution\ preserves comparable performance benefits in real-world situations and provides enhanced incremental deployment capabilities compared to PS, RAR, and other PS-based INA methodologies.

\section{Discussion}\label{sec:discussion}

% \subsubsection{Optimizing Downlink Utilization}

% In the \solution\ design, the ScatterReduce stage optimally leverages the uplink (from worker to INA switch), leaving the downlink (from INA switch to worker) partially idle. This provides an opportunity for improved utilization. With refined scheduling, \solution\ should be capable of fully mitigating the downlink traffic, thereby enabling parallel execution of the ScatterReduce and AllGather stages. Research on relevant synchronous scheduling schemes is currently ongoing.

\subsubsection{Improving Robustness}

\solution\ confers a centralized control mechanism, enhancing the robustness of the cluster. However, specific design details and comprehensive evaluations are still needed. We are currently conducting experiments to make \solution\ to provide increased reliability, expedited error recovery, and dynamic scalability, further strengthening the robustness of DDL training clusters. 
Moreover, \solution\ does not address the issue of node heterogeneity. The heterogeneity in node computational capabilities should be managed through other methods such as batch-size tuning~\cite{chen2023osp}.

\subsubsection{Combined with Model Parallelism}

Although this paper primarily discusses data parallelism and does not address model parallelism, it is worth noting that RAR is also frequently used in widely-used model parallel synchronization strategies. Consequently, if we attempt to introduce INA into model parallelism, \solution\ can be seamlessly integrated. We are continuing to investigate the potential challenges and solutions associated with incorporating \solution\ into model parallelism.

\section{Related Work}\label{sec:related_works}

\subsubsection{Reduce the Communication Size in DDL}

Communication compression approaches are proposed to reduce communication costs. Stochastic Rounding~\cite{gupta2015deep} randomly rounds the parameters in a method that preserves the expected value of the parameters. QSGD~\cite{alistarh2017qsgd} generalizes stochastic rounding to stochastic quantization and proposes multi-level gradient quantization schemes to further lower the transmission costs. 
% Gradient sparsification methods aim to reduce the amount of elements that are transmitted at each iteration. 
Gradient sparsification can also reduce the communication size. A representative method of gradient sparsification~\cite{strom2015scalable} uses a static threshold to decide which gradients to send. LTP~\cite{chen2023boosting} utilizes the loss-tolerant transmission to reduce the communication time. Deep gradient compression~\cite{lin2017deep} considers local gradient accumulation and guarantees the convergence of the training by accumulating momentum locally. These methods are orthogonal to \solution. However, all INA approaches require modifications to meet the needs of these methods.

\subsubsection{Communication Synchronization}

Synchronization models greatly affect the performance of DDL training. BSP~\cite{valiant1990bridging} is a classical synchronous framework. Stale-synchronous parallel (SSP)~\cite{ho2013more} aims to alleviate the straggler problem of BSP without losing synchronization by allowing faster workers to do more updates without waiting for slower ones, but still guarantees a staleness-bounded barrier. Compared with the SSP, ASP~\cite{lian2015asynchronous} eliminates the synchronization. Each work transmits its gradients to the PS after it calculates the gradients. OSP~\cite{chen2023osp} uses a 2-stage synchronization to reduce communication and speed up the training throughput. Local SGD~\cite{stich2018local} allows all workers to run a specific number of local updates independently before synchronization to guarantee good training accuracy. These approaches modified the synchronization architectures and are not applicable to INA capabilities.

\subsubsection{INA Approaches}

SwitchML~\cite{sapio2021scaling} design a communication primitive to perform parts of the model aggregation within the network. 
% Their evaluations are deployed on the Tofino platform and show an increase in training performance. 
ATP~\cite{lao2021atp} explores the idea of partitioning aggregation functionality between switches and servers so as to seamlessly support multi-tenant scenarios. PANAMA~\cite{gebara2021network} proposes a special transport layer protocol for load balance and congestion control. ASK~\cite{he2023generic} uses a key-value data structure for in-network aggregation to support compressed gradients aggregation. INAlloc~\cite{temp:inalloc} takes switch memory resources into consideration and designs a memory management mechanism to fully utilize memory resources in clustered switches. These methods are mostly discussed based on the PS scenario and have not attempted to integrate INA capabilities into RAR.

\section{Conclusion}\label{sec:conclusion}

In this study, we find that state-of-the-art PS-based INA approaches like ATP lack incremental deployment capabilities through mathematical modeling. This is detrimental to the construction and upgrade of the existing data center. Based on these issues, we propose \solution, which is known as the first to introduce INA capabilities into the RAR architecture. \solution\ not only provides excellent incremental deployment capabilities, but also greatly alleviates the problem of long dependency chain issues in RAR for throughput degradation. Through extensive testbed and simulation evaluations, we verify that \solution\ achieves better throughput and deployment cost-effectiveness than PS, RAR, and PS-based INA approaches under various topologies, models, and datasets.

\section*{Acknowledgement}

This work is sponsored by the Key-Area Research and Development Program of Guangdong Province (2021B0101400001), the National Natural Science Foundation of China (62172108), the Major Key Project of PCL, and the Natural Science Foundation of Shanghai (23ZR1404900). 

We sincerely appreciate the anonymous reviewers for their valuable and constructive feedback.

%% file: Contents/EvaluationFigures.tex
\begin{figure*}[tbp]
    \centering
    \includegraphics[width=0.6\textwidth]{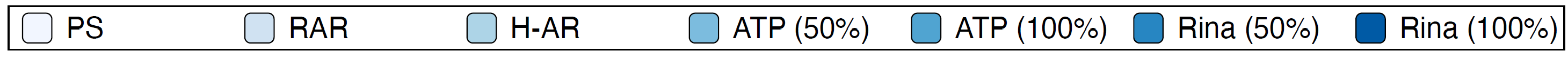}
    \\
    \subfigure[Fat-tree (k=4) with 16 workers.]{\includegraphics[width=0.46\textwidth]{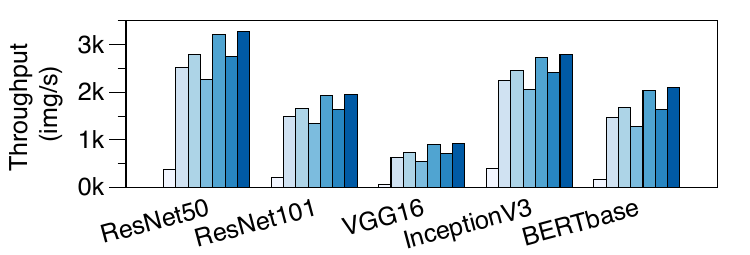}
    \label{fig:thru:fattree}}
    \subfigure[Dragonfly (a=4, g=9, and h=2) with 72 workers.]{\includegraphics[width=0.46\textwidth]{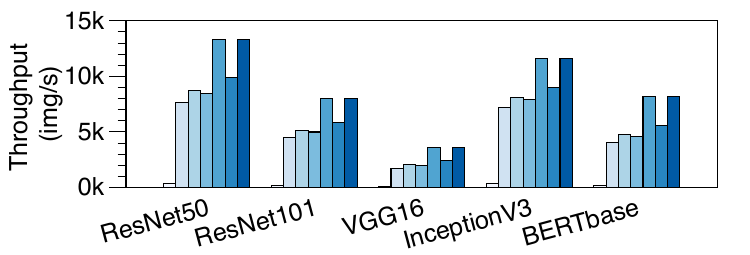}
    \label{fig:thru:df}}
    
    \caption{Evaluation on throughput through large-scale simulation.}
    \label{fig:evaluation:throughput}
    \vspace{-0.2in}
\end{figure*}